\documentstyle[12pt]{article}
\def\a{\alpha}\def\b{\beta}
\def\g{\gamma}\def\d{\delta}\def\e{\epsilon}

\def\l{\lambda}
\def\L{\Lambda}
\def\th{\theta}

\newcommand{\p}[1]{(\ref{#1})}
\begin{document}
\renewcommand{\thefootnote}{\fnsymbol{footnote}}
\begin{flushright}
Preprint DFPD 95/TH/46\\
hep-th/95xxx \\
September, 1995
\end{flushright}

\vspace{0.5cm}
\begin{center}
{\Large \bf Space--time
symmetries in duality symmetric models.\footnote{This work
was carried out as part of the
European Community Programme
``Gauge Theories, Applied Supersymmetry and
Quantum Gravity'' under contract SC1--CT92--D789, and
supported in part by M.P.I.}
}

\vspace{1cm}
{\bf Paolo Pasti},
\renewcommand{\thefootnote}{\dagger}
{\bf Dmitrij Sorokin\footnote{on leave from Kharkov Institute of
Physics and Technology, Kharkov, 310108, Ukraine.\\~~~~~e--mail:
sorokin@pd.infn.it} and
\renewcommand{\thefootnote}{\ddagger}
Mario Tonin\footnote{e--mail: tonin@pd.infn.it}
}

\vspace{0.5cm}
{\it Universit\`a Degli Studi Di Padova,
Dipartimento Di Fisica ``Galileo Galilei''\\
ed INFN, Sezione Di Padova,
Via F. Marzolo, 8, 35131 Padova, Italia}

\vspace{1.cm}
Talk given at the workshop on ``Gauge theories, applied
supersymmetry and quantum gravity''\\
(Leuven, Belgium, 10--14 July, 1995)

\vspace{1.cm}
{\bf Abstract}
\end{center}

\bigskip
A way of covariantizing duality symmetric actions is described.
As examples considered are a manifestly space--time invariant
duality--symmetric action for abelian gauge fields coupled to
axion--dilaton fields and gravity in D=4, and a Lorentz--invariant action
for chiral bosons in D=2. The latter is shown to admit
a manifestly supersymmetric generalization
for describing chiral superfields in n=(1,0) D=2 superspace.
\renewcommand{\thefootnote}{\arabic{footnote}}
\setcounter{footnote}0
\newpage

We would like to report  on a covariantization of
duality symmetric actions in various space--time dimensions considered
earlier by
\begin{quotation}
\noindent
Zwanziger (1971) for Maxwell fields in D=4 \cite{z};\\
Floreanini and Jackiw (1987) for chiral bosons in D=2
\cite{j};\\
Henneaux and Teitelboim (1987) for self--dual tensor fields
in $D=4p+2$ ($p$=1,2,...)\cite{ht};\\
Tseytlin (1990) for a duality symmetric string \cite{ts};\\
Schwarz and Sen (1993) for dual fields in any D
\cite{ss2,ss1}.
\end{quotation}

In connection with recent progress in understanding the important role
of duality symmetries in Yang--Mills theories \cite{mo,h,sw},
supergravity and string theory \cite{n}--\cite{sch},
the construction of versions of these
theories where duality would be a manifest symmetry of the action may
help to gain new insight into the structure of these theories.

A particular feature of duality symmetric models considered so far
\cite{j}--\cite{ss1}  is that the presence of duality symmetry violates
manifest Lorentz and general coordinate invariance and supersymmetry of
the action, conventional space--time symmetries being replaced by some
modified transformations of fields.

However, usually it is desirable to have a space--time covariant
formulation, which makes the structure of the theory more transparent
and often reveals new properties and new links between different parts
of the theory.

So, one can admit that these duality symmetric actions are, in
fact, a gauge choice in more general manifestly space--time and duality
symmetric models with richer local symmetry structure \cite{l,pst,pst1}.
We shall show that this is indeed the case by use of D=4 Maxwell theory
and a D=2 chiral field model as examples.

\section{Duality in Maxwell theory}
It is well known that the action for
a free Maxwell field $A_m(x)$ ($m$=0,1,2,3)
\begin{equation}\label{max}
S=-\int d^4x{1\over 4}F_{mn}F^{mn}={1\over 2}\int d^4x({\bf E}^2-{\bf
B}^2)
\end{equation}
is not invariant under duality transformations of the electric and
magnetic strength vectors ${\bf E}~\rightarrow~{\bf B},~{\bf
B}~\rightarrow~-{\bf E}$, while the free Maxwell equations are
invariant:
\begin{equation}\label{me}
\nabla\times{\bf E}=-{{\partial{\bf B}}\over{\partial t}},
\qquad
\nabla\times{\bf B}={{\partial{\bf E}}\over{\partial t}},
\qquad
\nabla{\bf E}=\nabla{\bf B}=0.
\end{equation}

To have duality symmetry at the level of action one has to double the
number of gauge fields ($A^\alpha_m$, $\alpha=1,2$) \cite{z,ko,ss2}
and construct an action in such a way that one of
the gauge fields becomes an auxiliary field upon solving equations of
motion \cite{ss2}. An alternative duality symmetric version of Maxwell
action was considered in \cite{d}. The duality symmetric action of
refs. \cite{z,ss2} can be written in the following form:
\begin{equation}\label{ss}
S=\int d^4x(-{1\over 8}F^\alpha_{mn}F^{mn\alpha}
+{1\over 4}{\cal F}^\alpha_{0i}{\cal F}^{\alpha}_{i0}),\qquad (i=1,2,3)
\end{equation}
where
\begin{equation}\label{sd}
{\cal F}^\alpha_{mn}={\cal L}^{\alpha\beta}F^\beta_{mn}-{1\over
2}\epsilon_{mnlp}F^{lp\alpha}={1\over 2}\epsilon_{mnlp}
{\cal F}^{lp\beta}{\cal L}^{\alpha\beta},
\end{equation}
(${\cal L}^{12}=-{\cal L}^{21}=1$).

Duality symmetry is a discrete subgroup of $SO(2)$ rotations of
$A^\alpha_m$
$(A^\alpha_m~\rightarrow~{\cal L}^{\alpha\beta}A^\beta_m)$.

Note that ${\cal F}^\alpha_{mn}{\cal F}^{\alpha mn}=0$
due to the self--duality
property, and the second term in \p{ss} breaks manifest Lorentz
invariance. However, beside the manifest spacial rotations the action
\p{ss} is invariant under the following modified space--time
transformations of $A^\alpha_i$ (in the gauge $A^\alpha_0=0$)
\begin{equation}\label{st}
\d A^\alpha_i=x^0v^k\partial_kA^\alpha_i+v^kx^k\partial_0A^\alpha_i+
v^kx^k{\cal L}^{\alpha\beta}{\cal F}^\beta_{0i},
\end{equation}
where the first two terms describe the ordinary Lorentz boosts along
a constant velocity $v^i$ and the third term vanishes
on the mass shell since an additional local symmetry of the action
\p{ss}
\begin{equation}\label{ad}
\d A^a_0=\varphi^\alpha(x)
\end{equation}
allows one to reduce the equations of motion
\begin{equation}\label{em}
{{\d S}\over{\d A^\alpha_i}}=\e^{ijk}\partial_i{\cal F}^\alpha_{k0}=0
\end{equation}
to a duality condition
\begin{equation}\label{9}
{\cal F}^\alpha_{mn}={\cal L}^{\alpha\beta}F^\beta_{mn}-{1\over
2}\epsilon_{mnlp}F^{lp\alpha}=0
\end{equation}
which, on the one hand, leads to the Maxwell equations
\begin{equation}\label{ma}
\partial_m{\cal F}^{mn\alpha}=\partial_mF^{mn\alpha}=0
\end{equation}
and, on the other hand, completely determines one of the gauge fields
through another one. For instance, using the relation
$$
{1\over 2}\epsilon_{mnlp}F^{2mn}=F^1_{mn}
$$
we can exclude $A^2_m(x)$ from the action \p{ss} and get the
conventional Maxwell action.

\bigskip
{\bf Covariantization}. One can admit that the action \p{ss}
arose as a result of some gauge fixing which specifies time direction
in a Lorentz covariant action \cite{pst,pst1}.
\footnote{An attempt to construct this action was
undertaken by Khoudeir and Pantoja \cite{l}. Unfortunately, they did not
manage to find a consistent formulation of the model.}

The first step is to covariantize the self--dual part of the
action \p{ss}. The simplest possible way to do this is to introduce an
auxiliary vector field $u_m(x)$ and write the action as follows:
\begin{equation}\label{11}
S_A=\int d^4x (-{1\over 8}F^\alpha_{mn}F^{\alpha mn}
+{1\over{4(-u_lu^l)}}u^m{\cal F}^\alpha_{mn}{\cal F}^{\alpha np}u_p).
\end{equation}

The main problem is to find a local symmetry which would permit to
choose a gauge where $u_m$ is a constant vector, in particular,
\begin{equation}\label{g}
u_m(x)=\d^0_m.
\end{equation}
Then the action \p{11} can reduce to \p{ss}. Note that we have
introduced the norm of $u_m$ $(u^lu_l=u^2)$ into the action \p{11}. This
is necessary for ensuring the symmetry we are looking
for. When $u_m$ is a constant vector, it corresponds to the frozen
straight Dirac string \cite{z}, while when $u_m(x)$ depends on
space--time coordinates, one can regard the Dirac string to be curved.

The search for this symmetry turns out to be connected with another
problem, namely, the problem of preserving a local symmetry under \p{ad}.
In the covariant version this transformation should be replaced by
\begin{equation}\label{adc}
\d A^\alpha_m=u_m\varphi^\alpha.
\end{equation}
To keep this symmetry is important (as we have already seen) for getting
the duality condition \p{9}.

To have the invariance under transformations \p{adc} one should add to
the Lorentz invariant action \p{11} another term
\begin{equation}\label{B}
S_B=-\int d^4x\e^{mnpq}u_m\partial_nB_{pq},
\end{equation}
where $B_{mn}(x)$ is an antisymmetric tensor field. Then the variation
of \p{11} under \p{adc} is canceled by the variation of \p{B} under
\begin{equation}\label{bv}
\d B_{mn}=-{{\varphi^\alpha}\over{u^2}}({\cal F}^\alpha_{mp}u^pu_n-
{\cal F}^\alpha_{np}u^pu_m).
\end{equation}
Note that \p{B} is also invariant under
\begin{equation}\label{b}
\delta B_{mn}=\partial_{[m}b_{n]}(x).
\end{equation}

As in the case of action \p{ss}, the local symmetry \p{adc} allows one to
fix a gauge on the mass shell in such a way that the duality condition
\p{9} takes place.

Thus, we again remain with only one independent Maxwell field and get
the duality between its electric and magnetic strength vector. In view of
the vanishing condition for the self--dual strength tensor the equations
of motion of $u_m$ reduce to:
\begin{equation}\label{u}
{{\d(S_A+S_B)}\over{\d u_m}}=\e^{mnlp}\partial_nB_{lp}=0 ~~\rightarrow~~
B_{mn}=\partial_{[m}b_{n]},
\end{equation}
which means that $B_{mn}$ is completely auxiliary and can be eliminated
by use of the corresponding local transformations \p{b}.

The only thing which has remained to show is that $u_m$ itself does not
carry physical degrees of freedom and can be gauge fixed to
$u_m=\d^0_m$. For this we have to find a corresponding local symmetry.
The form of the action \p{B} prompts that $S_A+S_B$ (\p{11}, \p{B})
can be invariant under the following transformations of $u_m$:
\begin{equation}\label{pc}
\d u_m(x)=\partial_m\varphi(x).
\end{equation}
This is indeed the case provided $A^\alpha_m$ and $B_{mn}$ transform as
follows
\begin{equation}\label{12}
\d A^\alpha_m=
{{\varphi(x)}\over{u^2}}{\cal L}^{\alpha\beta}{\cal F}^\beta_{mn}u^n,
\qquad
\d B_{mn}={{\varphi(x)}\over{(u^2)^2}}{\cal F}^{\alpha r}_{m}
u_r{\cal F}^{\beta s}_nu_s{\cal L}^{\alpha\beta}.
\end{equation}
Then, taking into account that the equations of motion of $B_{mn}$ give
\begin{equation}\label{ue}
\partial_{[m}u_{n]}=0~~~\rightarrow~~~u_m(x)=\partial_m\hat\varphi(x)
\end{equation}
and requiring that $u^2\not= 0$ (to escape singularities), we can use this
local transformation to put $u_m=\d^0_m$. In this gauge the manifestly
Lorentz invariant duality symmetric action
\begin{equation}\label{lo}
S=\int d^4x (-{1\over 8}F^\alpha_{mn}F^{\alpha mn}
+{1\over{4(-u_lu^l)}}u^m{\cal F}^\alpha_{mn}{\cal F}^{\alpha np}u_p
-\epsilon^{mnpq}u_m\partial_nB_{pq}).
\end{equation}
reduces to \p{ss}, and the local transformations of
$A_m^\alpha$ \p{12} (with $\varphi(x)=x^iv^i$) are combined with the
corresponding Lorentz transformations and produce the modified
space--time symmetry \p{st} of the action \p{ss}.

It seems of interest to understand the origin of the fields $u_m(x)$ and
$B_{mn}(x)$ and of the local transformations \p{pc}.

The form of the term in the action \p{lo} containing $B_{mn}$ reminds a
term one encounters in a formulation of a pseudoscalar  (`axion') field
(see, for the details \cite{o,bo,nt,k,q} and references therein)
\begin{equation}\label{1}
S=\int d^4x \left(
-{1\over 2}(\partial_m a(x)-u_m(x))(\partial^m a(x)-u^m(x))
-\e^{pqmn}u_p\partial_qB_{mn} \right).
\end{equation}

The action \p{1} is invariant under local Peccei--Quinn transformations
\begin{equation}\label{pct}
\d a(x)=\varphi(x),\qquad \d u_m=\partial_m\varphi(x),
\end{equation}
($u_m$ being the corresponding gauge field) and produces dual
versions of the axion action:
$$
L=-{1\over 2}\partial_ma(x)\partial^ma(x),
$$
\begin{equation}\label{a}
L={1\over{3!}}\partial_{[m}B_{np]}\partial^{[m}B^{np]}.
\end{equation}
The duality relation between the pseudoscalar field $a(x)$ and the
antisymmetric tensor field $B_{mn}$
\begin{equation}\label{5}
\partial_la(x)=\epsilon_{lmnp}\partial^mB^{np}
\end{equation}
is a consequence of the equations of motion of $u_m$ obtained from
\p{1}.

Thus, one may treat the origin of the duality symmetric Maxwell action
as a result of a specific coupling of the two Maxwell fields to the
auxiliary gauge field $u_m$ from the axionic sector of the theory.

\bigskip
{\bf $SL(2,{\bf R})$ invariant axion--dilaton coupling and coupling to
gravity}.\\
The duality symmetric model considered above admits
axion--dilaton coupling which respects global $SL(2,{\bf R})$ invariance
as well as coupling to gravity \cite{ss2,pst,pst1}.

To do this we introduce a $2\times 2$ matrix dilaton--axion field
\begin{equation}\label{calm}
{\cal M} =
{1\over \lambda_2(x)}\pmatrix{1 & \lambda_1(x)\cr \lambda_1(x) &
\lambda^2_1+\lambda^2_2\cr},
\qquad {\cal M}^T={\cal M}, \qquad {\cal{MLM}}^T={\cal L},
\end{equation}
and define the
global $SL(2,{\bf R})$ transformations
as follows:
\begin{equation}\label{om}
{\cal M}~~\rightarrow~~\omega^T{\cal M}\omega, \qquad \omega{\cal
L}\omega^T={\cal L}, \qquad A_m=\omega^TA_m.
\end{equation}

The coupling is carried out by the modification of the self--dual tensor
\p{sd} in the following way
\begin{equation}\label{sdm}
{\cal F}^{\alpha}_{mn}={\cal L}^{\alpha\beta}F^{\beta}_{mn}
-{1\over{2\sqrt{-g}}}\epsilon_{mnlp}
({\cal L}^T{\cal {ML}})^{\alpha\b}F^{lp\beta}
\equiv {\sqrt{-g}\over 2}
({\cal L}{\cal M})^{\alpha\beta}
\epsilon_{mnpq}{\cal F}^{\beta pq},
\end{equation}
where $g=\det g_{mn}(x)$ is the determinant of a gravitational field
metric.

In the most simple form the $SL(2,{\bf R})$ invariant action is written
as follows:
\begin{equation}\label{a2}
S=\int d^4x\sqrt{-g}(
{1\over {2u^2}}u^mF^{*\alpha,a}_{mn}
{\cal F}^{\alpha,a~ np}u_p
-{1\over 4}g^{mn}tr(\partial_m{\cal {ML}}\partial_n{\cal{ML}})
+{1\over{\sqrt{-g}}}\e^{pqmn}u_p\partial_qB_{mn}+R).
\end{equation}
Upon fixing the gauge
$u_m={1\over{\sqrt{-g^{00}}}}\delta^0_m$,
$B_{mn}=0$, the action \p{a2}
directly reduces to a corresponding Schwarz--Sen action \cite{ss2}.

Note that we would like to
identify the field $a(x)$ with $\l_1(x)$ from
the axion--dilaton matrix \p{calm},
then, due to the specific coupling
of the scalars to the abelian gauge fields the local Peccei--Quinn
transformations of the axion field are
broken down to the global shifts which
are part of the $SL(2,{\bf R})$ group. This is reflected in the
structure of the kinetic term for the
scalar fields in \p{a2} which does
not contain the auxiliary gauge field $u_m$. However, the action
is still invariant under the
local transformations of $u_m$, $A^\alpha_m$
and $B_{mn}$ \p{pc}, \p{12}.
One can go even further and eliminate from
the action \p{a2} the term containing $B_{mn}$ by substituting into
\p{a2} $u_m=\partial_m\hat\varphi(x)$.

The action \p{a2} can be extended to include $O(6,22)$ scalar fields
corresponding to a low energy bosonic sector of a toroidally
compactified heterotic string in a straightforward way \cite{ss2,pst1}.

\bigskip
{\bf Adding fermions and supersymmetry}.\\
The kinetic term for neutral
fermions is added to the duality symmetric action of a free Maxwell
field
without any problems:
\begin{equation}\label{sus}
S=\int d^4x(-{1\over 8}F^\alpha_{mn}F^{\alpha mn}
+{1\over{4(-u_lu^l)}}u^m{\cal F}^\a_{mn}{\cal F}^{\a np}u_p
-\e^{mnpq}u_m\partial_nB_{pq}-i\overline{\psi}\g^m\partial_m\psi).
\end{equation}
This action is supersymmetric
\cite{ss2,pst,pst1} \footnote{A superfield generalization of the
non--covariant action \p{ss} was
considered earlier in \cite{sal}.} under the
following transformations with odd constant parameters
 $\e^\a=i\g_5{\cal L}^{\a\b}\e^\b$:
$$
\d A^\a_m=i{\overline{\psi}}\g_m\e^\a,
$$
\begin{equation}\label{susy}
\d{\psi}={1\over 8}F^{\a mn}\g_m\g_n\e^\a-
{1\over{4u^2}}{\cal L}^{\a\b}u_p{\cal F}^{\a
pm}u^n\g_m\g_n\e^\b,
\end{equation}
all other fields being inert
under the supersymmetry transformations.

We see that the supersymmetric transformation law for the fermion
$\psi(x)$ (\ref{susy}) is non--conventional and
reduces to the ordinary
SUSY transformations of the vector supermultiplet $(A^1_m(x),~\psi(x))$
only on the mass shell (\ref{9}) upon excluding $A^2_m$:
\begin{equation}\label{su}
\d A^1_m=i\bar\psi\g_m\e^1;\qquad \d\psi={1\over 4}F^{1mn}\g_m\g_n\e^1.
\end{equation}
This is analogous to the problem with the
Lorentz transformations (\ref{st}) which we have just solved.
Using the same reasoning as lead us to
introducing $u_m(x)$ one may try
to find a superpartner of $u_m(x)$ whose
presence in the theory gives
rise to a local fermionic symmetry
(being a counterpart of (\ref{pc},
\ref{12})) which involves $\psi(x)$
and leads to (\ref{susy}) upon gauge
fixing the local fermionic symmetry.

We shall demonstrate the existence of such  local fermionic symmetry in
a simpler model for supersymmetric chiral fields in two space--time
dimensions.

\section{Chiral bosons and fermions in D=2}

For comparison let us consider a formulation of the dynamics of the
simplest possible self--dual field, namely a chiral boson $\phi(x)$,
in two--dimensional space--time. On the mass shell $\phi(x)$ satisfies a
self--duality (chirality) condition:
\begin{equation}\label{b1}
{\cal F}_m\equiv\partial_m\phi-\e_{mn}\partial^n\phi=0=
(\partial_0-\partial_1)\phi=\partial_{--}\phi
\end{equation}
and describes right--moving modes. ${\cal F}_m=\e_{mn}{\cal F}^n$, and
$(--,++)$ denote the light--cone vector components.

There are several (classically equivalent) versions of the chiral boson
action \cite{si,j,bar} from which the chirality condition \p{b1} is
obtained. A Lorentz covariant action proposed by Siegel \cite{si}
contains the square of the condition $\partial_{--}\phi=0$ with a
corresponding Lagrange multiplier $\lambda_{++++}(x)$:
\begin{equation}\label{b2}
S=\int d^2x{1\over 2}(\partial_{++}\phi\partial_{--}\phi-
\lambda_{++++}(\partial_{--}\phi)^2).
\end{equation}
The system described by the action \p{b2} turns out to be anomalous at
the quantum level (because of the constraint $(\partial_{--}\phi)^2=0$)
and requires an additional Wess--Zumino term to cancel the anomaly
\cite{im}.

By putting $\l_{++++}=1$, which breaks manifest Lorentz invariance, one
gets the Floreanini--Jackiw action \cite{j}
\begin{equation}\label{b3}
S=\int d^2x{1\over 2}(\partial_{++}\phi\partial_{--}\phi-
(\partial_{--}\phi)^2),
\end{equation}
which describes a quantum mechanically consistent chiral boson system
(provided, appropriate boundary conditions on $\phi$ are
imposed) \cite{j,g}, since the constraint
$(\partial_{--}\phi)^2=0$ does not directly follow from \p{b3}. The
ordinary Lorentz transformations of $\phi(x)$ are replaced by:
\begin{equation}\label{b31}
\d\phi=vx_m\e^{mn}\partial_n\phi-vx^1\partial_{--}\phi,
\end{equation}
where $v$ is a constant parameter, and the first term is the ordinary
Lorentz boost.
The sum of the actions \p{b3}, one for left--movers and one for
right--movers, describes a duality symmetric model being a ground for
duality symmetric formulation of bosonic string theory \cite{ts}.

As in the case of Maxwell theory we can restore the manifest Lorentz
invariance by introducing into \p{b3} an auxiliary vector field
$u_m(x)$. Because of peculiar properties of the two--dimensional model,
this can be carried out in two classically equivalent ways. The first
possibility is to consider $u_m(x)$ as a unit--norm time--like vector
and to write the action in the form
\begin{equation}\label{b4}
S=\int d^2x{1\over 2}(\partial_{++}\phi\partial_{--}\phi-
u^m{\cal F}_mu^n{\cal F}_n).
\end{equation}
Then, because $u_m$ was required to satisfy $u_mu^m=-1$, it contains
only one independent component, and one can show that \p{b4} reduces to
the Siegel action \p{b2}.

Another possibility, which seems to be more appropriate from the quantum
point of view, is to construct a Lorentz covariant action by analogy
with the action \p{lo}:
\begin{equation}\label{b5}
S=\int d^2x{1\over 2}\left(\partial_{++}\phi\partial_{--}\phi+{1\over u^2}
u^m{\cal F}_mu^n{\cal F}_n -\e^{mn}u_m\partial_nB\right),
\end{equation}
where $B(x)$ is an auxiliary scalar field.
The action \p{b5} is invariant under the following local
transformations:
\begin{equation}\label{b6}
\d u_m=\partial_m\varphi,\qquad \d\phi={\varphi\over{u^2}}u^m{\cal F}_m,
\qquad \d B=\varphi\left({{u^m{\cal F}_m}\over{u^2}}\right)^2,
\end{equation}
which permit to choose the gauge $u_m=\d^0_m$ on the mass shell, where:
\begin{equation}\label{b7}
{{\d S}\over{\d B}}=0 ~~\rightarrow~~u_m=\partial_m\hat\varphi(x),
\end{equation}
and reduce \p{b5} to \p{b3}.

One can simplify \p{b5} by substituting into the action the expression
\p{b7} for $u_m$. Then \p{b5} takes the following form:
\begin{equation}\label{b8}
S=\int d^2x{1\over 2}\left(\partial_{++}\phi\partial_{--}\phi-
{{\partial_{++}\hat\varphi}\over{\partial_{--}\hat\varphi}}
(\partial_{--}\phi)^2\right),
\end{equation}
and the transformations \p{b6} reduce to
\begin{equation}\label{b9}
\d\hat\varphi=\varphi(x),\qquad\d\phi={{\varphi(x)}\over
{\partial_{--}\hat\varphi}}\partial_{--}\phi.
\end{equation}

Note that in contrast to the Siegel case \p{b2} the variation of \p{b8}
with respect to the auxiliary field $\hat\varphi(x)$ does not produce
the anomalous constraint $(\partial_{--}\phi)^2=0$, since $\hat\varphi$
enters \p{b8} under the derivative.

Adding to \p{b5} or \p{b8} the analogous action
(containing the same field $u_m$) for left--movers and taking an
appropriate number of the left-- and right--moving scalar fields one
obtains the duality symmetric formulation of
a string \cite{ts,ss1} with manifest space--time symmetries.

The Lorentz invariant form \p{b8} of the chiral boson action is directly
generalized to a supersymmetric case.

Let us consider bosonic superfields
$\Phi(x,^{--},x^{++},\th^+)=\phi(x)$ + $i\th^+\psi_+(x)$ and
$\L(x^{--},x^{++},\th^+)$ = $\hat\varphi(x)+i\th^+\chi_+(x)$,
which obey the conventional transformation law under
global shifts $\d\th^+=\e^+$, $\d x^{++}=i\th^+\e^+$ and $\d x^{--}=0$ in
$n=(1,0)$ flat superspace. Then the superfield generalization of \p{b8}
is
\begin{equation}\label{b10}
S=\int d^2xd\th^+{1\over 2}\left(D_+\Phi\partial_{--}\Phi-
{{D_+\L}\over{\partial_{--}\L}}(\partial_{--}\Phi)^2\right)
\end{equation}
and that of \p{b9} is
\begin{equation}\label{b11}
\d\L=C(x^-,x^+,\th^+),\qquad \d\Phi=
{{C(x^-,x^+,\th^+)}\over{\partial_{--}\L}}\partial_{--}\Phi,
\end{equation}
where $D_+={\partial\over{\partial\th^+}}+i\th^+\partial_{++}$,
$D_+^2=i\partial_{++}$ is the supercovariant derivative.

Using the transformations \p{b11} one can gauge fix all components of
$\L(x,\th)$, except $\L|_{\th=0}=\hat\varphi(x)$, to zero and identify the
latter with the time coordinate $x^0={1\over 2}(x^{++}+x^{--})$. Then, upon
integrating over $\th^+$ one gets a component action (considered in
\cite{ali} as part of a duality invariant string action with
non--manifest supersymmetry):
\begin{equation}\label{b12}
S=\int d^2x\left({1\over 2}(\partial_{++}\phi\partial_{--}\phi-
(\partial_{--}\phi)^2) - i\psi_+\partial_{--}\psi_+\right),
\end{equation}
which is invariant under the modified Lorentz transformations for the
bosonic field $\phi(x)$ \p{b31} and modified supersymmetry
transformations:
\begin{equation}\label{b13}
\d\phi=i\e^+\psi_+;\qquad \d\psi=i\e^+\partial_{++}\phi-
i\e^+\partial_{--}\phi
\end{equation}
for the fermionic field $\psi_+$ analogous to \p{susy}.

{}From \p{b12} one gets that on the mass shell $\phi(x)$
and $\psi_+(x)$ are chiral:
$$
\partial_{--}\phi=0,~~~~~ \partial_{--}\psi_+=0.
$$

\section{Discussion}
The models considered above represent examples of duality symmetric
models with manifest space--time symmetries. We have seen that in D=4
duality between two abelian gauge fields arose as a result of their
coupling to the auxiliary vector field which can be treated as the
gauge field of local Peccei--Quinn symmetry. The structure of the action
admits space--time invariant $SL(2,{\bf R})$ axion--dilaton coupling,
coupling to gravity and supersymmetrization. The example of chiral
(self--dual) fields in two space--time dimensions demonstrates the
possibility of constructing duality symmetric models in the form which
respects both the conventional Lorentz invariance and conventional global
supersymmetry.

The generalization of the results obtained to duality symmetric actions
for abelian gauge fields in other dimensions \cite{j}--\cite{ss2} is
rather straightforward.

As far as supersymmetry is concerned, it would be of interest to
construct a superfield supergravity generalization of the duality
symmetric models. The field content of the bosonic sector considered
above points to the possible existence of such supergravity though
its consistent formulation may turn out to be not an easy problem.

Other challenging problems are the extension of the
class of duality symmetric actions with that describing non--abelian
gauge fields, and involving into the consideration charged matter.
The discussion of problems of coupling electrically and magnetically charged
matter fields to duality symmetric electro--magnetic fields the reader
may find in \cite{yu} and references therein.

\bigskip
{\bf Acknowledgements}. The authors would like to thank N. Berkovits,
P. Howe, C. Imbimbo, B. Nilsson, P. van Nieuwenhuizen,
C. Preitschopf, E. Sezgin and A. Tseytlin
for interest to this work and discussion.

\end{document}